\documentstyle[12pt]{article}

\begin{document}
\newcommand {\be}{\begin{equation}}
\newcommand {\ee}{\end{equation}}
\newcommand {\bea}{\begin{array}}
\newcommand {\cl}{\centerline}
\newcommand {\eea}{\end{array}}
\renewcommand {\thefootnote}{\fnsymbol{footnote}}
\baselineskip 0.65 cm
\begin{flushright}
IC/99/190\\
hep-th/9912100
\end{flushright}
\begin{center}
{\Large{\bf Closed String Brane-Like States, Brane Bound States 

and Noncommutative Branes}}
\vskip .5cm

D. Polyakov and  M.M. Sheikh-Jabbari
\footnote{ E-mail:polyakov,jabbari@ictp.trieste.it} \\

\vskip .5cm

 {\it The Abdus Salam International Centre for Theoretical Physics 

 Strada Costiera, 11

I-34014, Trieste, Italy}\\
\end{center}

\vskip 2cm
\begin{abstract}
We study the mass and different RR charge distributions of the BPS (p,p-2)-brane
bound states in the closed string brane-like $\sigma$-model.
We show that such brane bound states can be realized by introducing a
constant B field in the closed string theory. In addition we show that 
the worldvolume coordinates of these brane bound states turn out 
to be noncommutative.  

\end{abstract}
\newpage
\section{ Introduction}

It is believed that D-branes play a crucial role in the dynamics of type II
string theories at strong couplings. In the usual treatment, D-branes are 
described perturbatively by open strings satisfying
Dirichlet boundary conditions for the transverse directions \cite{POL}. 
In an alternative approach
the non-perturbative dynamics of D-branes in string theory is described in 
terms of exotic massless vertex operators, brane-like states. 
In the {\it open string sector} one of these operators is described by
the five form BRST invariant vertex\cite{DIM1}:
\be\bea{cc}
V_{m_1...m_5}^{(-3)}=\oint e^{-3\phi}\psi_{m_1}..\psi_{m_5}{dz\over 2\pi
i}+\; ghosts, 
\eea\ee
where $\psi_m$ are ten dimensional NSR worldsheet fermions, $\phi$ is 
bosonized superconformal ghost and we have skipped the purely ghost terms 
which play no role in correlation functions but which are in principle
necessary to insure the BRST invariance.   
The remarkable property of these vertex operators is
that they exist at non-zero ghost pictures only. These vertex operators also 
appear as central terms in the picture changed
space-time superalgebra in the NSR string theory, i.e.  due to the no-go
theorem they should be
related to the dynamics of extended objects, such as D-branes.   
In the {\it closed string sector} the corresponding 5-form BRST invariant
propagating brane-like vertex is given by\cite{DIM2}:
\be\bea{cc}\label{1}
V_{5}{(p)}=\lambda_5(p)\epsilon_{a_1..a_4}\int d^2z 
e^{-3\phi}\psi_{a_1}..\psi_{a_4}\psi_t \bar{\partial}X^t e^{ip_{a}X^{a}}.
\eea\ee
The space-time indices in the $V_5$ vertex are split in the $(4+6)$ way,
$a_1,..., a_4=0,1,2,3 \;\;\ t=4...9$.
It is crucial that the BRST invariance condition confines the propagation of 
this vertex to four longitudinal dimensions.
Moreover the condition of worldsheet conformal invariance restricts the 
momentum dependence of the four dimensional $\lambda_5$ field as:
\be
\lambda_5(p)={\lambda_0\over p^4},
\ee    
where $\lambda_0$ is some constant presumably related to mass or RR charge of 
the D3-brane. It is shown that the NSR sigma model with the $V_5$ vertex in 
{\it flat space-time} is equivalent to GS superstring theory in the 
background of the D3-brane, namely the $AdS_5\times S^5$ \cite{TSY}, 
and reproduces the correlation functions of the large N ${\cal
N}=4,\;D=4$ SYM theory\cite{DIM3}. The $V_5$ vertex is therefore presumed to 
play an
important geometrical role, accounting for, in particular, the dynamical
compactification of the flat ten dimensional space-time to another maximally
supersymmetric $AdS_5\times S^5$ vacuum. In the formula (\ref{1}) the 
longitudinal 
4-dimensional indices of the $V_5$ operator correspond to the worldvolume of
D3-brane or the boundary of $AdS_5$. 

Apart from individual branes one can realize the non-marginal bound states
of the branes, i.e. bound states of p, p-2, p-4, ...-branes or bound
states of D-branes with fundamental strings. In the 
open string theory language these bound states can be introduced by the
open strings with mixed boundary conditions at the ends parallel to 
branes and Dirichlet boundary conditions for the transverse coordinates
\cite{{AS},{S}}:
\be\left\{\bea{cc}
g_{\mu\nu}(\partial-\bar{\partial})X^{\nu}+B_{\mu\nu}(\partial+\bar{\partial})
X^{\nu}|_{z=\bar z}=0 \;\;\; \mu,\nu=0,..., p \\
(\partial+\bar{\partial})X^{\mu}=0  \;\; \mu=p+1,..,9.
\eea\right.
\ee
The electric mixing, $B_{0i}\neq 0$, leads to brane-string bound state
while the magnetic mixing, $B_{ij}\neq 0$, produces the brane-brane bound
states. Quantizing the open strings with the above boundary conditions
reveals the noncommutative structure of the brane worldvolume
coordinates \cite{{AAS},{Ho}}.

In this paper we will study the closed string $\sigma$-model with the
$V_5$
brane-like state in the presence of constant non-zero B field. In the
section 2, we will
show that the mass density of the D3-brane in the presence of the B field
is reproduced by the scattering amplitude (correlation function) of two
$V_5$ operators (that describe the D3-brane dynamics), which is in exact 
agreement with the well-known results. In order to elucidate the
corresponding D3-D1 bound state, we analyze the scattering of RR fields 
in the above mentioned $\sigma$-model. We find that the structure of the
relevant four point correlation functions, $<V_5V_5 V^p_{RR}V^{p'}_{RR}>$,     
shows that our $\sigma$-model has the correct RR charge distributions of
the corresponding D3-D1 bound state.
In the section 3, studying the two point $X^{a}\;X^{b}$ correlation
functions in the above {\it closed string} $\sigma$-model, we show that
the noncommutative structure of the brane bound state worldvolume appears   
just like in the open string case. Namely, we will argue that 
the regularization of the closed sting amplitudes with the $V_5$
insertions leads to the well-known open string propagators in the presence
of B field, however this result appears to be regularization dependent.  
In the concluding section we discuss possible connections between brane
like $\sigma$-models, and NCSYM theories and the related interesting
questions.  

\section{Brane bound states and brane-like states }   
   
In this section we study the partition function of the closed string
theory with inserted $V_5$ state and constant B field:

\be\label{action}
Z[\lambda_5, B]=\int DX e^{\int g_{\mu\nu}\partial X^{\mu}\bar{\partial}X^{\nu} 
+B_{\mu\nu}\partial X^{\mu}\bar{\partial}X^{\nu}+\lambda_5\;V_5+\lambda_5\;
{\bar V}_5}
\ee
In order to explore the resulting brane structure we study the partition
function perturbatively in $\lambda_0$ and B field.
\be\label{Z}
Z=\sum_{n=0}^{\infty}{1\over n!}<V_5 V_5 V_B^n>+ O(\lambda^3),
\ee
where $V_B=B_{\mu\nu}\partial X^{\mu}\bar{\partial}X^{\nu}$ and we use the 
usual  OPE's for closed strings.
The first non-trivial contribution to the above sum comes from $V_B^2$,
which in the proper ghost picture is
\be\bea{cc}
<V_5 V_5 V_B V_B>=\int d^4k{\lambda_0^2\over k^4}\epsilon_{a_1..a_4}
\epsilon_{b_1..b_4} B_{i_1j_1} B_{i_2j_2}
<e^{\phi-{\bar\phi}}\psi_{a_1}..\psi_{a_4}\psi_{t_1} \bar{\psi}X^{t_1} 
e^{ik_{a}X^{a}}(z_1,\bar z_1)\\
e^{-3\phi-{\bar\phi}}\psi_{a_1}..\psi_{a_4}\psi_{t_2} \bar{\psi}^{t_2}
e^{-ik_{a}X^{a}}(z_2,\bar z_2)
\partial X^{i_1}\bar{\partial}X^{j_1}(z_3, \bar z_3)
\partial X^{i_2}\bar{\partial}X^{j_2}(z_4, \bar z_4)>.
\eea\ee

Since the B field  polarization is  parallel to the brane and it is an
antisymmetric tensor, there are no contractions between the $V_5$
and $V_B$ operators and hence we can easily find the above correlation
function
\be
<V_5 V_5 V_B V_B>=\lambda_0^2 det(g) Tr(B^2),
\ee
and for the same reason we can find all the contributions to (\ref{Z}).
Summing up all of
those terms, up to second order in ${\lambda_0}$ we find
\be\bea{cc}
Z[\lambda_5,B]=Z[\lambda_5,B=0]{det{(g+B)}\over det{g}}\\
\;\;\;\;\;\;\;=Z[\lambda_5 \sqrt{{det{(g+B)}\over det{g}}},B=0].
\eea
\ee\label{mass}

As we discussed in the introduction $\lambda_5$ shows the RR charge and
hence the mass density of the BPS D3-brane. In the same manner the
(\ref{mass}) shows that upon introducing the B field, the BPS mass
density of
the related extended object is given by, ${\rm D3-mass}\times
\sqrt{det{(g+B)}}$.

Since the constant B field term in the (\ref{action}) is  quadratic 
in $X$'s, the above calculations can be understood and performed by
replacing $g_{ij}$ by $g_{ij}+B_{ij}$ in  the usual closed string
propagators (both in bosonic and fermionic parts), i.e.
\be
  :X_i(z_1)X_j(z_2):|_{B}=(g_{ij}+B_{ij})log|z_1-z_2|.
\ee
Hence $Z=<V_5V_5>|_{B}+O(\lambda^3)$, which clearly reproduces the above
results.

To show that turning on the B-field in the
brane-like  $\sigma$-model (5) leads to the bound D3-D1 state,
one has to study the scattering of probe  RR-fields off
the $V_5$-state (which corresponds to D3-brane in our model)
in the presence of the B-term. The bound D3-D1 state
(its appearance is the effect of switching on the B-field)
manifests itself through the presence of the appropriate
Ramond-Ramond charge distributions in the system. The relevant correlation
function is
given by:
\be\bea{cc}
A^{V_5-RR}
(z_1,{\bar{z_1}},...z_4,{\bar{z_4}})=<{V^{(1,0)}_5}(z_1,{\bar{z_1}})
{V^{(-3,0)}_5}(z_2,{\bar{z_2}}){V_{RR}^{(-{1\over2},-{1\over2})}}
(z_3,{\bar{z_3}}){V_{RR}^{({1\over2},-{3\over2})}}(z_4,\bar{z_4})>\\
{V^{(1,0)}_5}(z_1,{\bar{z_1}},k_1^{||})=e^{\phi}\psi_0\psi_1\psi_2\psi_3\psi_{t_1}
\bar\partial{X^{t_1}}e^{ik_1^{||}X}(z_1,\bar{z_1})\\
{V^{(-3,0)}_5}(z_1,{\bar{z_1}},k_2^{||})=e^{-3\phi}\psi_0\psi_1\psi_2\psi_3\psi_{t_2}
\bar\partial{X^{t_2}}e^{ik_2^{||}X}(z_2,\bar{z_2})\\
V_{RR}(z_3,{\bar{z_3}})=e^{-{\phi\over2}-{{\bar\phi}\over2}}
\Sigma_{\alpha_1}{\Gamma^{m_1...m_5}_{\alpha_1\beta_1}}
\bar\Sigma_{\beta_1}e^{ip_1X}(z_3,{\bar{z_3}})F^{RR}_{m_1...m_5}(p_1)\\
V_{RR}(z_4,{\bar z}_4)=e^{{\phi\over2}-{{3\over2}{\bar\phi}}}
\Sigma_{\gamma_2}C\;\Gamma^m_{\gamma_2\alpha_2}\partial X_m
{\Gamma^{n_1...n_3}_{\alpha_2\beta_2}}
\bar\Sigma_{\beta_1}e^{ip_2X}(z_4,{\bar{z_4}})F^{RR}_{n_1...n_3}(p_2)
\eea
\ee
The upper indices in brackets refer to the left and right
ghost numbers of vertex operators, $\Sigma$ are space-time spin operators,
the momenta $k_1$, $k_2$ of the $V_5$-vertices are four-dimensional
(polarized along the 0,1,2,3 longitudinal directions), while
the momenta of the 3-form and 5-form RR-vertices 
are ten-dimensional.
The presence of the constant B-field results in the modification of the
O.P.E. between spin  fields and fermions.
First of all, the two-point function of the 
longitudinal worldsheet fermions (parallel to the D3-brane worldvolume)
is given by:
\be
<\psi^{a}(z)\psi^b(w)>={{\eta^{ab}+B^{ab}}\over{z-w}}
\ee
while the correlators of the transverse fermions are unchanged.
For simplicity, we will consider the B-field with the only non-vanishing
$B_{23}$ component elsewhere in this paper.
Performing the usual construction of space-time spinors out of
worldsheet fermions with taking into account the modification
related to the B-field, we find that the relevant modified O.P.E.s
 are given by:

\be\left\{\bea{cc}
\Sigma_\alpha(z)\Sigma_\beta(w)\sim{{{{\tilde\epsilon}_{\alpha\beta}}\over
{(z-w)^{5\over4}}}}+\sum_p{{{\tilde\Gamma^{m_1...m_p}_{\alpha\beta}}
\psi_{m_1}...\psi_{m_p}}\over{(z-w)^{{5\over4}-p}}}\\
\Sigma_\alpha(z)\psi^b(w)\sim
{{{\tilde\Gamma^b_{\alpha\beta}}\Sigma_\beta}\over{(z-w)^{1\over2}}}+...\\
{\tilde\Gamma^{m_1...m_p}}\equiv{\Gamma^{m_1...m_p}}+
{\Gamma^{23}}B_{23}\Gamma^{m_1...m_p}\\
{\tilde\epsilon}_{\alpha\beta}\equiv\epsilon_{\alpha\beta}+
{\Gamma^{23}_{\alpha\beta}}B_{23}
\eea\right.
\ee
The part of the correlation function, consisting of holomorphic
ghosts, fermions and spin operators is given by:
\be\bea{cc}
<e^\phi\psi_0...\psi_3\psi_{t_1}(z_1)e^{-3\phi}\psi_0...\psi_3(z_2)
e^{-{\phi\over2}}\Sigma_{\alpha_1}(z_3)e^{\phi\over2}\Sigma_{\alpha_2}(z_4)>\\
={{\eta^{t_1t_2}{\tilde\epsilon_{\alpha_1\alpha_2}}(z_1-z_4)}\over
{(z_1-z_2)^2(z_1-z_3)(z_3-z_4)}}+
{{\Gamma_{\alpha_1\alpha_2}^{t_1t_2}}\over{(z_1-z_2)(z_1-z_3)^2}}\\
+{{{({\tilde\Gamma}^{0123t_1}{\tilde\Gamma}^{0123t_2})_{\alpha_1\alpha_2}}
(z_1-z_4)}\over{(z_1-z_3)^2(z_1-z_2)(z_2-z_4)}}+
{{{({\tilde\Gamma}^{012}{\tilde\Gamma^{012}})_{\alpha_1\alpha_2}}\eta^{t_1t_2}
+permut.(0,1,2,3)}\over{(z_1-z_3)(z_3-z_4)(z_1-z_2)^2}}\\
+{{{({\tilde\Gamma}^{01t_1}{\tilde\Gamma}^{01t_2})_{\alpha_1\alpha_2}
+permut.(0,1,2,3)}\over{(z_1-z_3)(z_1-z_2)^2}}}+
{{{({\tilde\Gamma}^{0123t_1}{\tilde\Gamma}^{0123t_2})_{\alpha_1\alpha_2}
(z_1-z_2)(z_1-z_3)^3}\over{(z_1-z_4)^3(z_2-z_3)^3}}}\\
{{{({\tilde\Gamma}^{012}{\tilde\Gamma^{012}})_{\alpha_1\alpha_2}}\eta^{t_1t_2}
+permut.(0,1,2,3)}\over{(z_1-z_4)^2(z_2-z_3)^3(z_1-z_2)^{-1}(z_1-z_3)^{-1}}}\\
+{{{({\tilde\Gamma}^{01t_1}{\tilde\Gamma}^{01t_2})_{\alpha_1\alpha_2}
+permut.(0,1,2,3)}
\over{(z_1-z_4)^2(z_1-z_2)^{-1}(z_1-z_3)^{-1}(z_2-z_3)^3}}}\\
+{{\eta^{t_1t_2}\delta_{\alpha_1\alpha_2}
+(\Gamma^{t_1}\Gamma^{t_2})_{\alpha_1\alpha_2}}\over{(z_1-z_4)(z_2-z_3)^2}}\\
\eea
\ee
The antiholomorphic fermionic part of the correlator is given by:
\be
<e^{-{{\bar\phi}\over2}}{\bar\Sigma}_{\beta_1}(\bar{z_3})
e^{-{{3\bar\phi}\over2}}{\bar\Sigma}_{\gamma_2}(\bar{z_4})>
={{{\tilde\epsilon}_{\beta_1\gamma_2}}\over{(\bar{z_3}-\bar{z_4})^2}}
\ee
The evaluation of the bosonic X-dependent part gives:
\be\bea{cc}
<\bar\partial{X^{t_1}}e^{ik_1X}(z_1,\bar{z_1})\bar\partial{X^{t_2}}
e^{ik_2X}(z_2,\bar{z_2})e^{ip_1X}(z_3,\bar{z_3})
\partial{X_m}e^{ip_2X}(z_4,\bar{z_4})>\\=
|z_1-z_2|^{-2\lbrace{k_1k_2}\rbrace}|z_1-z_3|^{-2\lbrace{k_1p_1}\rbrace}
|z_1-z_4|^{-2\lbrace{k_1p_2}\rbrace}\\
|z_2-z_3|^{-2\lbrace{k_2p_1}\rbrace}|z_2-z_4|^{-2\lbrace{k_2p_2}\rbrace}
|z_3-z_4|^{-2\lbrace{p_1p_2}\rbrace}\\
\lbrace({{\eta^{t_1t_2}}\over{(\bar{z_1}-\bar{z_2})^2}}+
{{p_1^{t_1}p_1^{t_2}}\over{(\bar{z_1}-\bar{z_3})(\bar{z_2}-\bar{z_3})}}+
{{p_2^{t_1}p_2^{t_2}}\over{(\bar{z_1}-\bar{z_4})(\bar{z_2}-\bar{z_4})}}
+{{p_1^{t_1}p_2^{t_2}}\over{(\bar{z_1}-\bar{z_3})(\bar{z_2}-\bar{z_4})}}
+{{p_2^{t_1}p_1^{t_2}}\over{(\bar{z_1}-\bar{z_4})(\bar{z_2}-\bar{z_3})}})\\
({{\tilde{p_1^m}}\over{z_3-z_4}}+{{\tilde{k_2^m}}\over{z_2-z_4}}+
{{\tilde{k_1^m}}\over{z_1-z_4}})\rbrace
\eea
\ee
The ``B-covariant'' momentum ${\tilde{p^m}}$ is related to the
usual one as
\be\left\{\bea{cc}
{\tilde{p^2}}=p^2+B_{23}p^3\\
{\tilde{p^3}}=p^3-B_{23}p^2\\
{\tilde{p^m}}\equiv{p^m},m=0,1,4,...,9.
\eea\right.
\ee
whereas the ``B-covariant'' scalar product is defined as
\be
\lbrace{p_1p_2}\rbrace=(p_1)_m(p_2)^m+B_{23}(p_1^2p_2^3-p_1^3p_2^2)
\ee 

Finally, putting together all the pieces of the correlator
$A^{V_5-RR}(z_1,{\bar{z_1}},...z_4,{\bar{z_4}})$,
fixing the SL(2,C) remnant conformal symmetry by setting
$z_2\rightarrow{1},z_{3}\rightarrow{0},z_{4}\rightarrow\infty$
and integrating over $z_1$ with the appropriate Koba-Nielsen's 
measure, we obtain the following expression for the scattering
amplitude:
\be\left\{\bea{cc}A^{V_5-RR}(k_1,k_2,p_1,p_2)=\int{d^2}{z_1}
|z_2-z_3|^2|z_2-z_4|^2|z_3-z_4|^2
A^{V_5-RR}(z_1,{\bar{z_1}},...z_4,{\bar{z_4}})\\=
3(Tr(C\Gamma^{n_1...n_3}\Gamma^m\Gamma^{m_1...m_5})+
Tr(C\Gamma^{n_1...n_3}\Gamma^2\Gamma^3\Gamma^m\Gamma^{m_1...m_5})
B_{23})\\
\times\int{d^2}{z_1}|1-z_1|^{-2(\lbrace{k_1k_2}\rbrace+2)}
|z_1|^{-2\lbrace{k_1p_1}\rbrace}
\eea\right.
\ee
It is remarkable that, after fixing the global
SL(2,C) conformal invariance the only terms that survive in
the amplitude are those linear in $B_{23}$  (in the kinematic factor).
All the contributions to the kinematic factor which are non-linear in 
the B-field, go away as we take the limit $z_4\rightarrow\infty$.
Performing the integration over $z_1$, we get:
\be\bea{cc}A^{V_5-RR}(k_1,k_2,p_1,p_2)=3
\lambda_5(k_1)\lambda_5(k_2)F_{m_1...m_5}(p_1)
F_{n_1...n_3}(p_2)\\
(Tr(C\Gamma^{n_1...n_3}\Gamma^m\Gamma^{m_1...m_5})+
Tr(C\Gamma^{n_1...n_3}\Gamma^2\Gamma^3\Gamma^m\Gamma^{m_1...m_5})
B_{23}){(p_1)_m}\\
\times{{\Gamma(1+{1\over2}\lbrace{k_1k_2}\rbrace)
\Gamma(1-{1\over2}\lbrace{k_1p_1}\rbrace)
 \Gamma({1\over2}\lbrace{k_1(k_2+p_1)}\rbrace)}\over
{\Gamma(1-{1\over2}\lbrace{k_1(k_2+p_1)}\rbrace)
\Gamma(1-{1\over2}\lbrace{k_1k_2}\rbrace)
\Gamma({1\over2}\lbrace{k_1p_1}\rbrace)}}
\eea
\ee
This  expression has yet to be integrated  over the four-dimensional
momenta $k_1$ and $k_2$ of the $V_5$-vertices.
The integration similar to the one done in \cite{DIM3}, taking
into account the momentum conservation $p_2=-p_1-k_1-k_2$ along with
the condition (3), gives:

\be\bea{cc} 
A=\int{{{d^4k_1}}}\int{{d^4k_2}}
A^{V_5-RR}(k_1,k_2,p_1,p_2)=
3(\lambda_0)^2{F_{m_1...m_5}}(p_1)F_{n_1...n_3}(p_2)\\
{(p_1)_m}(Tr(C\Gamma^{n_1...n_3}\Gamma^m\Gamma^{m_1...m_5})+
Tr(C\Gamma^{n_1...n_3}\Gamma^2\Gamma^3\Gamma^m\Gamma^{m_1...m_5})
B_{23})log(p_1^{||})^2+...
\eea
\ee
where 
$p_1^{||}$ is the projection of the ten-dimensional momentum
of the RR-vertex to the four longitudinal dimensions, and 
we have skipped the contributions analytic in $p_1^{||}$
(which do not play any role upon the Fourier transformation
back to the position space).
Finally, taking traces of the gamma-matrices, we find that the
amplitude is proportional to:
\be
A\sim{{\lambda_0^2}F_{0123t}F^{01t}B_{23}}
\ee
corresponding to  the attraction force between the D3-brane and D-string
in the $D3-D1$ bound state.

Analogously, one can consider turning on the $B_{01}$ component
of the B-field, along with $B_{23}$.
In this case,the effect of the B-field in the $D3$-brane worldvolume
will result in appearance of the D-instanton-D-string-D3-brane  bound
state, in addition to the D1-D3 one. The related string-theoretic 
$V^{RR-V_5}$ scattering amplitude will be quadratic in the B-field, being
proportional to $A\sim\lambda_0^2{F_t}{F^{0123t}}B_{01}B_{23}$,
where $F_t$ is the RR field strength, corresponding to the 
D-instanton. Analogously, one can observe the appearance of other
various D-brane bound states as a result of switching on the B-field , 
in the context of the brane-like $\sigma$-model.

\section{Noncommutativity from closed string $\sigma$-model}

In this section we will consider the propagators of two bosonic closed string 
$X$ fields in the presence of the B field and the $V_5$ operator.  The
correlation function of interest is given by  
$$
<\partial X^a(z_1)\bar{\partial} X^b({\bar z}_2)>|_{B,V_5}.
$$ 
As it follows from the $\sigma$-model action (\ref{action}) that the first
non-trivial contribution to this correlator (of order of $\lambda_5^2$) 
is proportional to
\be\bea{cc}
<\partial X^a(z_1)\bar{\partial} X^b({\bar z}_2)>|_{B,V_5}\sim \lambda_0^2
\int { d^4k\over k^4} \int d^2w_1 d^2w_2 \\
<\partial X^m(z_1)\bar{\partial} X^n({\bar z}_2)
V_5^{(1,0)}(w_1,{\bar w}_1,k ) V_5^{(-3,0)}(w_2,{\bar w}_2,-k)>.
\eea\ee
The integration over $k$ is proportional to $\sim \int  {d^4k\over k^4}k^a k^b$
which gives $ g^{ab}$ after the appropriate regularization. The above
correlation function is then  
given by
\be\bea{cc}
<\partial X^a(z_1)\bar{\partial} X^b({\bar z}_2)>|_{B,V_5}\sim \lambda_0^2
\biggl( (g+B)g^{-1}(g+B) \biggr)^{ab} 
\int {d^2w_1 d^2w_2 \over |w_1-w_2|^4}\times\\
\{ {1\over (z_1-w_1)({\bar z}_2-{\bar w}_1)}+ {1\over (z_1-w_2)({\bar z}_2-{\bar 
w}_2)}- {1\over (z_1-w_1)({\bar z}_2-{\bar w}_2)}- 
{1\over (z_1-w_2)({\bar z}_2-{\bar w}_1)} \}. 
\eea\ee
Let us consider the first integral given by
\be
I(z_1,{\bar z}_2)=\int {d^2w_1 d^2w_2 \over |w_1-w_2|^4}\times 
{1\over(z_1-w_1)({\bar z}_2-{\bar w}_1)}.
\ee
Writing $w_1=x+iy$, we have
\be
I(z_1,{\bar z}_2)=\int d^2w_2 \int_{-\infty}^{\infty} dy  \int_{-\infty}^{\infty} dx
{1\over (x+iy-w_2)^2 (x+iy-{\bar w}_2)^2 (z_1-x-iy)({\bar z}_2-x+iy)}.
\ee
For the sake of certainty let us consider the case $Im(z_1), Im(z_2)\geq 0$, so
that both of $z_1$ and $z_2$ are in the upper half plane.   
Now in order to elucidate the open string structure of the propagator we 
propose to consider not all the poles in the above integral, but only those 
located in the upper half plane and then to retain only the terms
surviving the translational invariance along the $x$ axis. 
Evaluating residues of those poles in $x$ and then in $y$ we find 
\be
I(z_1,{\bar z}_2)=2\partial_{z_1} \partial_{{\bar z}_2}\int d^2w 
{1\over(z_1-w_2)({\bar z}_2-{\bar w}_2)}.
\ee
Our procedure for distinguishing the upper half plane poles contains few 
subtleties; see below for more explicit explanation.
Again we can perform the $w_2$ integration by writing it as $w_2=u+iv$, so that
the integral can be written as 
\be
I(z_1,{\bar z}_2)=2\partial_{z_1} \partial_{{\bar z}_2}
\int_{-\infty}^{\infty} dv  \int_{-\infty}^{\infty} du
{1\over (z_1-u-iv)({\bar z}_2-u+iv)}.
\ee
The essential point in our computation is that we only take into account the 
poles located in the upper half plane (this may be achieved e.g. by
including the exponential factor $\sim e^{i\alpha x}, \alpha>0$ in the
complex integral over $x$ 
and taking the limit $\alpha\rightarrow 0$ afterwards).  
\be
I(z_1,{\bar z}_2)=2\partial_{z_1} \partial_{{\bar z}_2}
\int_{0}^{\infty} dv {1\over ({\bar z}_2-z_1+2iv)}.
\ee
Now we shall stress that in the integration over $v$, the integration must be taken
from zero to $\infty$, rather than from $-\infty$ 
to $\infty$, in order to exclude the terms breaking  the translational
invariance in the direction of the real axis.
Finally, regularizing the above integral
and differentiating over $z_1$ and ${\bar z}_2$ we find
\be
I(z_1,{\bar z}_2)={1\over (z_1-{\bar z}_2)^2}.
\ee
The corresponding contribution to the $X\;X$ propagator is given by
$$(g+B)g^{-1}(g+B)log(z_1-{\bar z}_2)$$.  
Adding the necessary complex conjugate part coming from 
$$
<\bar{\partial} X^a(z_1){\partial} X^b({\bar z}_2)>|_{B,V_5}= 
(g-B)g^{-1}(g-B){1\over ({\bar z}_1-z_2)^2},$$ 
we get the corresponding part of the $X\;X$ propagator:
\be
\biggl((g+B)g^{-1}(g+B)\biggr)_{\bf S}log|z_1-{\bar z}_2|^2 + 
\biggl((g+B)g^{-1}(g+B)\biggr)_{\bf A}log{z_1-{\bar z}_2\over {\bar z}_1-z_2}
\ee
where the ${\bf S},{\bf A}$ show the symmetric and antisymmetric parts of that
matrix. 
To find the full expression for the propagator we also have to take into 
account the contributions coming from  
$<{\partial} X^a(z_1){\partial} X^b({\bar z}_2)>|_{B,V_5}$ and 
$<\bar{\partial} X^a(z_1)\bar{\partial} X^b({\bar z}_2)>|_{B,V_5}$. 
Adding all of these contributions together up to second order in
$\lambda_0^2$ we have
\be\bea{cc}\label{XX}
<X^a(z_1)\;X^b(z_2)>\sim -g^{ab}(log|z_1-z_2|-log|z_1-{\bar z}_2|)\\
\hspace{2 cm}             -(G^{ab}+\theta^{ab})log(z_1-{\bar z_2})-
                          (G^{ab}-\theta^{ab})log({\bar z_1}-z_2),
\eea\ee                            
where $G$ and $\theta$ are the the symmetric and antisymmetric parts of 
$(g+B)_{ab}^{-1}$. This is in agreement with the expression for open string 
$X\;X$ propagator in the constant B field background giving rise to the
noncommutative structure of the brane worldvolume coordinates \cite{SW}.       
       
In order to make the full correspondence with the open string picture,
we should also work out the closed string propagators for the transverse
coordinates and show that they result in the open string propagators with 
Dirichlet boundary conditions. To do this we have to take into account the
transverse degrees of freedom of the D3-brane which in the closed string 
language correspond to the vertex operators \cite{DIM2}
$$V_3={\lambda_0 \over (p_tp^t)^3}\epsilon_{a_1..a_4}\int d^2z \partial\{
e^{\phi}\psi_{a_1}..\psi_{a_3}\bar{\partial}X_{a_4} e^{ip_{t}X^{t}}\}+\;
ghosts,$$
where $p_t$ is the transverse momentum. The relevant correlator is given
by \newline
$<X^t(z,{\bar z}) X^{t'}(z',{\bar z}')>|_{V_5,V_3,B}$. Again instead
of the $X$'s we consider the correlators of their derivatives first. Up to
the second order in $\lambda_0$ the relevant correlation functions are
given by 
$<\partial X^t(z,{\bar z}) \bar{\partial} X^{t'}(z',{\bar z}')V_3 V_3>$
and
$<\partial X^t(z,{\bar z}) {\partial} X^{t'}(z',{\bar z}')V_5 V_5>$
and their complex conjugate which, as previously, should be integrated
over the coordinates and the momenta of the brane-like vertices.
The computation totally similar to the one performed above gives the
result
\be
<X^t(z,{\bar z}) X^{t'}(z',{\bar z}')>|_{V_5, V_3, B}\sim
\lambda_0^2 det(g+B) (log|z-z'|{\bf -}log|z-{\bar z}'|),
\ee
which coincides with the well-known expression for the open string
propagator with Dirichlet boundary conditions.

\section{Discussions and Remarks}

In this paper we have studied the adding of a constant B field background
to the closed string
brane-like $\sigma$-model. We argued that since the B term is quadratic in
$X$ and only contains the parallel $X$ components, the contractions
between  $V_B$ and brane-like states vanish and therefore the constant B field 
effects can be totally absorbed in the closed string propagators.
Calculating the mass density and different RR charge densities we
have shown
that turning on the B field in the brane-like $\sigma$-model gives rise to
the non-marginal D3-D1 bound state. In addition we have shown that the
closed string propagators in the presence of B field 
reveal the noncommutativity of the brane coordinates. All the
above results can be checked and clarified in the brane-brane
scattering processes \cite{{AS},{S}}. In order to study such scatterings
we should be able to identify the location of $D_p$-brane in the $9-p$
dimensional transverse space. This can be done by adding a delta function
of the brane location, $Y^t$, to the $V_5$ state:
$$
V_{5}(k_{||},k_t)=({\lambda_0\over k^4_{||}})\epsilon_{a_1..a_4}\int d^2z 
e^{-3\phi}\psi_{a_1}..\psi_{a_4}\psi_t \bar{\partial}X^t
e^{ik_{||}X^{||}}\;e^{ik_{t}(X^{t}-Y^{t})}.
$$     
We should note that adding such a term will not destroy the superconformal 
and the BRST invariance. 

As it has been argued in \cite{{DIM2},{DIM3}}, adding the brane-like
states to the closed strings in the flat background leads to strings
effectively living in the related supergravity backgrounds, 
namely $AdS_5\times S^5$. 
So we expect that the same ideology should work 
for the case we have at hand, the brane bound states. However, in
the presence of the B field, the related supergravity solution is not
$AdS$ anymore. It has been argued in \cite{{MR},{AOS}} that one can find a
limit in which the corresponding NCSYM theory (described in terms of 
properly scaled parameters) decouples from the bulk gravity.  
So it seems that we should be able to build some more concrete relations 
between string theory  in the above
mentioned brane bound state backgrounds and the deformed gauge theories on the
boundary.
In particular one can find the four point function of NCSYM theory, 
$<F^2_{NC}F^2_{NC}>$ from the dilaton scattering off these bound states.

The more interesting question we can address here is the Wilson loops.
Since it is believed that the $V_5$ operator (accounting for dynamical
compactification $AdS_5\times S^5$)  compensates
for the zig-zag non-invariance of the gravity part and since the B term is 
zig-zag invariant, Wilson loops of the NCSYM theory calculated by means of
the gravity/NCSYM correspondence should be zig-zag invariant.
 
It is well-known that the conformal invariance
should be broken in the confining phase. In the gauge theory language that
means that we are just living apart from the UV fixed point. From the
string theory point of view breaking the conformal symmetry 
in the four
dimensional space-time usually corresponds to turning on the gravity
in transverse dimensions (e.g. the radial coordinate of $AdS_5$)
\cite{{VV}}. However the presence of the transverse graviton modes 
is expected to break the zig-zag symmetry, necessary to relate the string
theory with the gauge-theoretic large N limit.
The alternative approach, which preserves the zig-zag symmetry, 
is based on including the axionic B term to the functional for the Wilson
loop. In this case one can expect this functional both to obey the
large N loop equation and to exhibit the area law behaviour, necessary for
the confinement.

{\bf Acknowledgements}

This research was partly supported by the EC contract no. ERBFMRX-CT
96-0090.

\end{document}